\documentclass[aps,prl,twocolumn,showpacs,superscriptaddress,10pt]{revtex4}
\usepackage{graphicx}
\usepackage{amsmath}
\usepackage{amsfonts}
\usepackage{amssymb}
\usepackage{amsthm}
\usepackage{times}
\usepackage[colorlinks,citecolor=blue,linkcolor=blue]{hyperref}
\usepackage{color}

\begin{document}
\title{Anomalous Heat Diffusion}

\newcommand{\NUS}{\affiliation{Department of Physics and Centre for Computational Science
and Engineering, National University of Singapore, 117546 Singapore, Singapore}}
\newcommand{\NGS}{\affiliation{NUS Graduate School for Integrative Sciences and
Engineering, 117456 Singapore, Singapore}}
\newcommand{\Augsburg}{\affiliation{Institute of Physics, University of Augsburg,
Universit\"atsstrasse 1, D-86159 Augsburg, Germany}}
\newcommand{\Tongji}{\affiliation{Center for Phononics and Thermal Energy Science,
School of Physics Science and Engineering, Tongji University, 200092 Shanghai, China}}
\newcommand{\USA}{\affiliation{Theoretical Division, Los Alamos National Laboratory, Los
Alamos, 87545 New Mexico, USA}}
\newcommand{\Munich}{\affiliation{Nanosystems Initiative Munich,
Schellingstr, 4, D-80799 M\"{u}nchen, Germany}}

\author{Sha Liu}
\email{phylius@nus.edu.sg}
\NUS
\NGS
\author{Peter H\"{a}nggi}
\email{hanggi@physik.uni-augsburg.de}
\NUS
\Augsburg
\Munich
\Tongji
\author{Nianbei Li}
\Tongji
\author{Jie Ren}
\USA
\author{Baowen Li}
\email{phylibw@nus.edu.sg}
\NUS
\NGS
\Tongji

\date{17 Dec 2013}

\newcommand{\expk}{\alpha}
\newcommand{\expx}{\beta}
\newcommand{\Sec}[1]{{\bf #1}}
\newcommand{\levy}{L\'evy}

\newcommand{\mean}[1]{\left\langle#1\right\rangle}
\newcommand{\dv}[1]{\delta\!\mean{#1}_{neq}}   
\newcommand{\non}[1]{\mean{#1}_{neq}}  
\newcommand{\eqi}[1]{\mean{#1}_{eq}}        

\newcommand{\NN}{\mathcal{N}}
\newcommand{\define}{\equiv}
\newcommand{\msdp}{\langle x_p^2(t)\rangle}
\newcommand{\msd}{\langle \Delta x^2(t)\rangle_E}
\newcommand{\Corr}{\mathcal{C}}

\newcommand{\textemph}[1]{\textcolor{blue}{\bf #1}}

\begin{abstract}
Consider anomalous energy spread in solid phases, i.e., $\msd \equiv
\int (x -{\langle x \rangle}_E)^2
\rho_E(x,t)dx \propto t^{\beta}$, as induced by a small initial   excess energy
perturbation distribution $\rho_{E}(x,t=0)$ away from equilibrium. The
second derivative
of this  variance of the nonequilibrium  excess energy distribution is shown to obey rigorously the intriguing relation, $d^2\msd/dt^2 = 2\Corr_{JJ}(t)/(k_BT^2c)$, where $\Corr_{JJ}(t)$ equals the
thermal equilibrium total heat flux  autocorrelation function and  $c$
is the specific volumetric  heat capacity.  Its integral assumes a {\it time-local}
Helfand-like relation.
Given that the  averaged nonequilibrium
heat flux is governed by an anomalous heat conductivity, {the energy
diffusion scaling
determines} a corresponding anomalous thermal conductivity scaling behavior.
\end{abstract}

\pacs{05.60.-k, 05.20.-y, 44.10.+i, 66.70.-f}

\maketitle

Fourier's law of heat conduction states the relation between local
heat flux and local temperature. In one dimension it
assumes the familiar form,  $j(x,t)=-\kappa \partial_x T(x,t)$,
where $j(x,t)$ is the local
heat flux density, $T(x,t)$ denotes the local equilibrium temperature and $\kappa$ is the
(normal) thermal conductivity.
Upon combining it with the {{local}} energy conservation
law $\partial_t
\varepsilon(x,t)+\partial_x j(x,t)=0$ and a local energy distribution relation
$\varepsilon(x,t)=c T(x,t)$, we arrive at the heat equation describing the normal spread
of energy,  reading
$\partial_t \varepsilon (x,t)=D_E \partial_{x}^2 \varepsilon(x,t)$,
wherein $c$ denotes the specific volumetric heat capacity and
$D_E=\kappa/c$ is the thermal diffusivity \cite{Helfand.60.PR}.

{{Although Fourier's law  is obeyed ubiquitously
in everyday experimental measurements for three-dimensional
 bulk materials possessing an inherent  anharmonicity, it nevertheless
remains an empirical law lacking a fundamental proof}}
\cite{Bonetto.00.NULL,Livi.03.Nature,Lepri.03.PR,Dhar.08.AP}. { An
open issue is its
validity in presence of spatial constraints caused by dimensionality.
Indeed, a longstanding, mainly theoretical debate over the last two
decades  indicates that  the Fourier law may fail in one- and two-dimensional momentum conserving systems; thus giving rise to
anomalous heat transport}
\cite{Lepri.97.PRL,Yang.06.PRE,Lepri.03.PR,Dhar.08.AP,Spohn.13.A}. In
such systems,
given a temperature bias $\Delta T$ across a sample of
length $L$, the nonequilibrium average heat flux  typically scales not
inversely with $L$, but instead obeys a length-dependent  scaling
relation, i.e.,
\begin{equation}
\label{eq:kappa}
J=\sigma(L) \Delta T \equiv \kappa(L) \frac{\Delta
T}{L}
\end{equation}
Here, $\sigma(L)$ denotes the heat conductance.
Commonly one  then formally
introduces $\kappa(L)\equiv\sigma(L) L$ as an effective heat
conductivity, which exhibits an
anomalous {\it length dependence} \cite{Lepri.03.PR,Dhar.08.AP}.
Therefore, a strictly  intensive material specific property as
heat conductivity generally does not exist; practically  at least not
on finite length scale.
A power law divergence $\kappa(L)\sim L^\alpha$, ($\alpha\neq 0$), is
typically observed for momentum conserving 1D systems, while for two
dimensional (2D)
systems $\kappa(L)\sim \log L$
\cite{Lepri.03.PR,Dhar.08.AP}.
It should be kept in mind however, that such an
effective thermal conductivity $\kappa(L)$ then generally does  {\it
not}  relate to the local heat flux density
in terms of a local temperature gradient; consequently, Fourier's
law in its usual form does no longer hold.

This intriguing length dependent behavior has not only
inspired  a vivid theoretical activity
\cite{Maruyama.02.PBCM,Zhang.05.JCP,Henry.08.PRL,Henry.09.PRB,
Nika.09.APL,Yang.10.NT,
Lindsay.10.PRB, Lindsay.11.PRB,Nika.12.NL,Liu.12.PRB} but as well several intriguing
recent experimental
investigations \cite{Chang.08.PRL,Hsiao.13.NN,Chang.13.MRS} on low
dimensional materials such as polyethylene
chains, single-walled carbon nanotubes and,
more generally, low-dimensional molecular chains.
In all these theoretical and experimental studies an anomalous
length
dependence for $\kappa(L)$  is clearly observed over extended
length ranges. --- Here, our main objective is how such a
length-dependent thermal conductivity behavior can uniquely be related
to inherent, anomalous diffusive energy spread in solid phases.

Because Fourier's law is connected to normal energy diffusion (see
above),
this violation of Fourier's law has been  studied as well from the
viewpoint of unbounded anomalous particle diffusion $x_p(t)$ in 1D
billiard models
\cite{Li.02.PRL,Alonso.02.PRE,Li.03.PRE,Denisov.03.PRL}, obeying
$\msdp\propto t^{\expx},\expx\ne1$. There, non-interacting particles
diffuse and transport ({\it kinetic}) energy anomalously. A scaling
relation $\expk=\expx-1$ was predicted
for the billiard models following
a \levy\ walk dynamics \cite{Denisov.03.PRL,Dhar.13.PRE}.
Notably, such a relation was verified by several numerical
investigations on energy diffusion in 1D lattice
systems
\cite{Cipriani.05.PRL,Zhao.06.PRL,Li.10.PRL,Zaburdaev.11.PRL,
Zaburdaev.12.PRL}.

Explicit analytical studies are, however,  available for
non-interacting \levy\ walk models only
\cite{Denisov.03.PRL,Dhar.13.PRE}.
Therefore, the result $\alpha=\beta-1$ is still
restricted to cases with non-confined particle diffusion rather than
with energy
diffusion in solid phases. With the particles executing small displacements about fixed lattice
sites, the energy transport in solids thus proceeds distinctly
different from
unconfined particle motion.
Put differently, the definition of  a mean square deviation (MSD) of
energy, i.e.  $\msd=
\mean{x^2(t)}_E-\mean{x}_E^2$,
along space $x$  has no direct meaning from an unconfined, diffusing
particle dynamics viewpoint. As a consequence, although those
previous efforts aimed at bridging  energy diffusion and heat conduction from the
viewpoint of particle diffusion are inspiring,
the general scheme of nonequilibrium energy diffusion
still remains an open issue.

Here, we study the general features of energy diffusion using linear
response theory. We derive  the evolution of the nonequilibrium excess energy density profile during
energy diffusion processes
\cite{Cipriani.05.PRL,Zhao.06.PRL,Zaburdaev.11.PRL,Zaburdaev.12.PRL}.
Based on this, we derive a dynamical equality which relates
the acceleration of nonequilibrium energy spread $\msd$ to the
equilibrium autocorrelation function of total heat flux
$\Corr_{JJ}(t)$.
This relation thus provides a sound and useful concept to investigate nonequilibrium,  generally anomalous heat
diffusion.

{\it Local excess energy distribution.}
In the following, { we limit the study of  energy diffusion
to isolated 1D systems with no energy and  particle exchange with
heat baths.}
The generalization  to higher dimensional cases is straightforward.

Typically, the diffusion of energy refers to a
relaxation process that an initially nonequilibrium energy
distribution evolves towards  equilibrium, just as the
relaxation of particle distribution in
normal diffusion. We
term this nonequilibrium distribution the excess energy
distribution, which is proportional to the
deviation
\cite{Cipriani.05.PRL,Zhao.06.PRL,Li.10.PRL,Zaburdaev.11.PRL,Zaburdaev.12.PRL},
$\dv{h(x,t)}\equiv \non{h(x,t)}-\eqi{h(x)}$, where
$\non{\cdot}$ denotes the expectation value in the nonequilibrium diffusion process,
$\eqi{\cdot}$ the equilibrium average and $h(x,t)$ the local
Hamiltonian
density. An illustration of this relaxation process is depicted with
Fig. \ref{fig:diff}(a) and (b) for the
relaxation of an arbitrarily chosen initial excess energy distribution
along an Fermi-Pasta-Ulam (FPU) chain
\cite{Berman.05.C,Dimensionless.00.NULL}.

{ Note that for  isolated, energy conserving systems this
total excess energy, $\delta
E=\int\dv{h(x,t)}dx$,  remains conserved \cite{Supplementary.00.NULL}.}
Therefore, the normalized fraction of excess energy at a certain position $x$
at time $t$ reads
\begin{equation}
	\label{eq:rho_def}
	\rho_E(x,t)= \frac{\dv{h(x,t)}}{\delta E}=\frac{\dv{h(x,t)}}{\int\dv{h(x,0)}dx}.
\end{equation}
This quantity formally presents the analog of a probability density
for particle diffusion. In distinct contrast, however, being a
reference density, it can take on negative values, cf. in Fig.
\ref{fig:diff}(a). Although not being a manifest probability density
it
nevertheless remains normalized
during time evolution, i.e.,
$\int \rho_E(x,t)dx=1$.
The MSD for energy diffusion thus reads
\begin{equation}
\label{eq:msd}
	\msd\!\equiv\!\! \int\! (x\!-\!{\langle x \rangle}_E)^2\! \rho_E(x,t)dx\!=\!
\mean{x^2(t)}_E\!-\!\mean{x}_E^2.
\end{equation}
Here, its first mean, ${\langle x \rangle}_E = \int x \rho_E(x,t)dx$, remains
constant in time, cf. in supplementary material
\cite{Supplementary.00.NULL}.
This  MSD $\msd$
can also assume    transient negative values;
reflecting the fact that it is  the  variance $\msd$ for this
nonequilibrium
excess energy distribution that spreads in time $t$ rather than the
equilibrium average $\langle(x(t)- x(t_0))^2\rangle_{eq}$ of the
displacements of particle positions \cite{Helfand.60.PR}.

A first main objective  { is  the evaluation of this very excess
energy
distribution
$\rho_E(x,t)$. In
doing so, we use  (Kubo)-linear response theory as put forward originally for an ensemble of isolated systems \cite{Kubo.57.JPSJ,Luttinger.64.PR,Zwanzig.65.ARPC,Visscher.74.PRA,
Allen.93.PRB}. We prepare at the infinite past a nonequilibrium state $f_{neq}$ in terms of a quenched canonical ensemble at temperature $T$, $f_{neq}\propto \exp(-\beta_{T} H_T)$, with a total Hamiltonian $H_T= H+H'$,  where $\beta_T=1/(k_BT)$, $H=\int h(x)dx$. Here, the part $H'$ accounts for the  applied small perturbation to $H$ by substituting in $H_T$ the local Hamiltonian density by $h(x)\rightarrow [h(x) -\eta(x)h(x)], \eta(x) \ll 1 $. This perturbation is then switched off suddenly  at time $t=0$ \cite{Supplementary.00.NULL}.
This so quenched initial nonequilibrium state subsequently undergoes an ergodic, isolated nonequlibrium dynamics governed by   the unperturbed
Liouvillian containing $h(x)$ only, which relaxes in the long time limit towards  the manifest  equilibrium
statistics with the canonical  phase space density $f_{eq}\propto \exp(-\beta_T
H)$.

As detailed in the supplementary material
\cite{Supplementary.00.NULL},
 the corresponding response function is given in terms of the equilibrium spatio-temporal correlation
of local Hamiltonian density $h(x,t)$. The result  explicitly reads
\begin{equation}
\label{eq:continuous}
	\dv{h(x,t)}= \frac{1}{k_BT}\int \Corr_{hh}(x,t;x',0)\eta(x')dx' \;,
\end{equation}
where for any two local quantities $a(x)$ and $b(x)$, we define
$\Corr_{ab}(x,t;x',t')\equiv {\langle \Delta a(x,t)\Delta b(x',t')\rangle} _{eq}$ with
$\Delta a(x,t)=
a(x,t)-{\langle a(x)\rangle}_{eq}$. Being in equilibrium}, these
spatial-temporal correlations obey
time-translational invariance,
i.e. $\Corr_{ab}(x,t+s;x',t'+s)=\Corr_{ab}(x,t;x',t')$ for arbitrary
$s$. For a homogeneous system, these
equilibrium correlations
$\Corr_{ab}(x,t;x',t')$ become spatially translation invariant,
yielding $\Corr_{ab}(x-x', t-t')$. Note that this requirement for
homogeneity does not exclude disordered situations; -- tailored disordered systems are also
homogeneous as long as the disorder strength is uniform. Consequently, the  total excess energy $\delta
E=\int\delta \langle h(x,0)\rangle_{neq} d{x}$ can be simplified to read
\begin{equation}
\delta E
=\!\!\iint dxdx' \Corr_{hh}(x\!-\!x',0)\frac{\eta(x')}{k_BT}\!= c T \!\int
\eta(x') dx',
\label{eq:cv}
\end{equation}
where $c$ is the volumetric specific heat capacity and $\int\Corr_{hh}(x,0)d{x}=k_B T^2 c$
has been used \cite{Supplementary.00.NULL}.
The normalized excess energy distribution (\ref{eq:rho_def}) then
reads
\begin{equation}
	\label{eq:rho}
	\rho_E(x,t)= \frac{1}{\NN}\int \Corr_{hh}(x-x',t)\eta(x') dx'\;,
\end{equation}
where $\NN=k_B T^2 c \int \eta(x) dx$ is the normalization constant.

{ For the nonequilibrium heat flow response it was not necessary to
make use of the concept of a spatially dependent temperature $T(x)$.
Such a spatially dependent temperature $T(x)$, if  indeed it exists,  would enter the result via the initial preparation of
the  quenched, displaced  thermal equilibrium upon   identifying the  quasi-force
$\eta(x)\equiv \delta T(x)/T  \ll 1$.  The energy distribution $h(x)$ then couples
formally to the conjugate thermodynamic affinity
$\delta T(x)/T$, implying that $\beta_T[1 - \delta T(x)/T]h(x) = \beta_T(x)h(x)$, cf. in
Refs.
\cite{Zwanzig.65.ARPC,Visscher.74.PRA,Allen.93.PRB}. Moreover, no
time-dependent
local equilibrium temperature $T(x,t)$  enters the derivation in  (\ref{eq:continuous}).}

\begin{figure}[t]
\includegraphics[width=1\columnwidth]{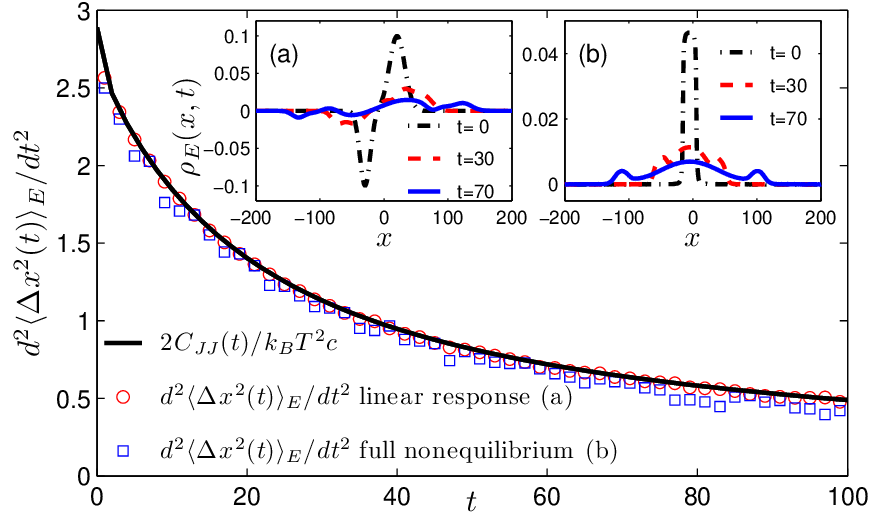}
\caption{(Color online) Numerical validation of the main result in \eqref{eq:main}
for an FPU chain with a
length $N=401$, specific heat $c=0.828$ at a dimensionless temperature $T=1$ \cite{Dimensionless.00.NULL}.  The red circles and the
blue squares are the second derivative $d^2\msd/dt^2$ as obtained from
the insets (a) and (b), respectively. The black solid line depicts
the result for the total heat flux autocorrelation
$\Corr_{JJ}(t)$,
i.e. the rhs in \eqref{eq:main}.
Insets: (a) energy
diffusion along the FPU chain using the linear response result
\eqref{eq:rho} with an initial small Hamiltonian perturbation $\eta(x)$ that is composed of two Gaussians of
opposite weights; (b) nonequilibrium energy
diffusion as obtained  from an initial  near equilibrium steady state.
For further details
see in \cite{Supplementary.00.NULL}.}
\label{fig:diff}
\end{figure}

{\it Anomalous energy diffusion vs. equilibrium heat flux correlation.}
The main result relating arbitrary  ergodic energy diffusion to the
equilibrium heat flux
autocorrelation function can  be obtained as follows:
With the conservation of local energy  $\partial_t
h(x,t)+\partial_x j(x,t)=0$, we obtain \cite{Supplementary.00.NULL}
\begin{equation}
\label{eq:hh-jj}
	\partial_t^2 \Corr_{hh}(x,t)=
	\partial_x^2 \Corr_{jj}(x,t).
\end{equation}
Additionally, define $J_L= \int_{-L/2}^{L/2} j(x,t)dx$ to be the total
heat flux for a 1D system of
length $L$, we have
\begin{equation}
\label{eq:ct}
    \Corr_{JJ}(t)\equiv\lim_{L\to\infty}\frac{1}{L}\mean{J_L(t)J_L(0)}_{eq}=
        \int_{-\infty}^{\infty}\Corr_{jj}(x,t)dx,
\end{equation}
This autocorrelation function of total heat flux $\Corr_{JJ}$ is the
central quantity that
knowingly
enters the Green--Kubo formula  for normal heat conductivity
\cite{Green.54.JCP,Kubo.57.JPSJ,Luttinger.64.PR,Zwanzig.65.ARPC,Visscher.74.PRA,
Allen.93.PRB}.

Upon combining Eqs. (\ref{eq:msd}), (\ref{eq:rho}), (\ref{eq:hh-jj}) and (\ref{eq:ct}), we
obtain the central result for the MSD:
\begin{equation}
\label{eq:main}
\begin{split}
	\frac{d^2\msd}{dt^2}&\!=\!\frac{1}{\NN}\!\!\iint\!\! x^2 \frac{\partial^2
\Corr_{hh}(x-x',t)}{\partial t^2}{\eta(x')} dxdx' \\
&\!=\!\frac{2\Corr_{JJ}(t)}{k_B T^2 c},
\end{split}
\end{equation}
where integration by parts has been used twice. This central equality constitutes an
equation of motion for the MSD of general energy diffusion. The corresponding
initial conditions are: ${\langle \Delta x^2(t=0)\rangle}_E=\iint
x^2\Corr_{hh}(x-x',0)\eta(x')d{x}d{x'}/\NN- \big(\iint
x\Corr_{hh}(x-x',0)\eta(x')d{x}d{x'}/\NN\big)^2$
and ${d \msd/dt}|_{t=0}= 0$.
It is only the initial value for $\msd$ that  exhibits a dependence on the initially
chosen  energy perturbation.
{The vanishing initial speed follows from the fact
that for an inertial dynamics $\Corr_{jj}(y,t)$ is  an even
function in time $t$, being continuously differentiable at time $t=0$.
Therefore, any physically realistic energy diffusion
process will start out  as  ballistic transport \cite{Huang.11.NP}.}

The  numerical verification of the main finding  in \eqref{eq:main}  is depicted  in Fig.
\ref{fig:diff}  for the theoretical archetype model of low-dimensional
heat transfer, i.e. for an FPU chain, as detailed in
\cite{Supplementary.00.NULL}.
Inset (a) is obtained
by evaluating the linear response result \eqref{eq:rho} at dimensionless $T=1$ from
an initial small perturbation  $\eta(x)$ with a positive and a negative Gaussian weight.
In inset (b), the full nonequilibrium energy diffusion is
simulated from an initial,
near equilibrium steady state using a preparation with heat baths of
differing temperature. The
energy diffusion proceeds after removing those heat baths. An ensemble of $4\times
10^8$ realizations
are used to obtain the depicted nonequilibrium energy density distribution
$\rho_E (x,t)$ in Fig. \ref{fig:diff}(b). The total heat flux
autocorrelation function $\Corr_{JJ}(t)$ is obtained in thermal
equilibrium at a temperature $T=1$ by averaging over an
ensemble of $2\times 10^9$ realizations. The specific heat, $c=0.828$,
is calculated analytically according to its
definition.
Very good agreement between theory and numerical experiments is obtained.

Let us  recall  the assumptions used in the  derivation of
this intriguing result: For the application of linear response theory the process is supposed to be sufficiently
ergodic, implying  that no nonstationary (i.e. aging) phenomena for long-time correlations are  at
work, thus ensuring  manifest relaxation  towards thermal
equilibrium. This crucial  ergodicity assumption rules out
all anomalous energy  diffusion processes that undergo aging, as  it
occurs
in many continuous time random walk descriptions
\cite{Metzler.00.PR,Barkai.03.PRL,Klafter.11.NULL,Klafter.11.NULLa,Sokolov.12.SM}. Those models, however, lack a microscopic
Hamiltonian basis. There exists, however, {\it ergodic} anomalous diffusion dynamics
stemming from  a Generalized Langevin equation  (GLE)
\cite{Goychuk.07.PRL,Kubo.66.RPP,Deng.09.PRE,Goychuk.09.PRE,
Goychuk.11.NULL,Siegle.11.EL,
Magdziarz.11.APNY,Sokolov.12.SM}. Likewise, microscopic Hamiltonian
models involving
homogeneous disordered lattices exhibit subdiffusive heat
conductivity \cite{Dhar.01.PRL,Roy.08.PRE}. Our result (\ref{eq:main}) is robust against changes in the initial energy
profile; it only affects the initial value of $\msd$. The  main
finding is restricted,
however, to near equilibrium situations; matters may change drastically  with  perturbations of the
system taken far away into nonequilibrium.

{\it Relation to the  Helfand scenario}.
Inspired by the Green-Kubo relation
\cite{Green.54.JCP,Kubo.57.JPSJ} for {\it normal} transport, Helfand
 showed that the  average over the canonical  initial thermal equilibrium of all phase
space coordinates of the squared displacement of the appropriate
 ``Helfand moment'', i.e.,  $G_L(t)=\int_{-L/2}^{L/2} x
\big(h(x,t)-\mean{h(x)}_{eq}\big)dx$,  obeys
$\langle[G_L(t)-G_L(0)]^2\rangle_{eq}/L= 2 \int_0^t (t-u) C_{JJ}(u)du$ \cite{Helfand.60.PR,Viscardy.07.JCP,Gaspard.08.JSMTaE}.
Therefore, taking the second time-derivative it follows with $L
\rightarrow \infty$, that
\begin{equation}
\label{eq:helfand}
	\lim_{L\to\infty}\!\frac{d^2}{dt^2}\frac{\langle[G_L(t)-G_L(0)]^2\rangle_{eq}}{L}
\!\equiv\!
	\frac{d^2\langle\Delta \mathcal{G}^2(t)\rangle_{eq}}{dt^2}
\!=\!2\Corr_{ JJ} (t).
\end{equation}
Here, the initial conditions are
$\langle\Delta \mathcal{G}^2(t=0)\rangle_{eq}=0$ and
$d\langle\Delta \mathcal{G}^2(t)\rangle_{eq}/dt|_{t=0}=0$.
Consequently, the scaled equilibrium average of the squared displacement of the Helfand moment, i.e.,   $\langle\Delta
\mathcal{G}^2(t)\rangle_{eq}/k_BT^2 c$, differs from $\msd$ by a constant shift, as
determined by the initially chosen excess energy profile.
In the absence of the main
relation
in (\ref{eq:main}), the mere result in (\ref{eq:helfand}) (with
dimension  $[\mathrm{length(energy)^2}]$) alone cannot provide the
result for the  spread $\msd$ of (anomalous) nonequilibrium energy diffusion. Observing the stated
initial conditions, we next integrate (\ref{eq:main}) to yield the
corollary
\begin{equation}
\label{eq:local_helfand}
	\frac{d\msd}{dt}= \int_0^t
\frac{2\ \Corr_{JJ}(t')}{k_BT^2c}dt'
=\frac{1}{k_BT^2 c} \frac{d\langle\Delta
\mathcal{G}^2(t)\rangle_{eq}}{ dt}.
\end{equation}
This finding can be interpreted as a {\it time-local}  Helfand-like
relation. This is so because in contrast to the ordinary Helfand
relation for
normal heat conductivity, i.e., $\kappa^{normal} = \langle\Delta
\mathcal{G}^2(t)\rangle_{eq}/(2t k_B T^2)$, no explicit time
derivative enters
\cite{Helfand.60.PR,Viscardy.07.JCP,Gaspard.08.JSMTaE}. Put differently,
(\ref{eq:local_helfand}) involves the time-local quantity $d\msd/dt$ (or $d\langle\Delta
\mathcal{G}^2(t)\rangle_{eq}/dt$) rather than a finite time version
$\msd/t$ (or
$\langle\Delta \mathcal{G}^2(t)\rangle_{eq}/t$). This intriguing  corollary
 (\ref{eq:local_helfand}) { assumes an appealing form} to establish
the
relationship between anomalous energy diffusion scaling and a  generally anomalous scaling  for the
thermal conductivity $\kappa(L)$ obeying $J \sim  \kappa(L) \Delta T/L$.

{\it Normal energy diffusion.}
For normal energy diffusion  the MSD  increases asymptotically
linearly in time, i.e.,
$\lim_{t\to\infty}\msd/t = 2 D_E$.
 $D_E$ is  termed the thermal diffusivity.
With time $t\to\infty$ in
(\ref{eq:local_helfand}) we find
\begin{equation}
\label{eq:kD}
\kappa^{normal}\!=\!\!\!\int_0^{\infty}\!\frac{\Corr_{JJ}(t)}{k_BT^2}d{t}=\!\frac{c}{2}
\!\lim_ {t\to\infty}\!\!\frac{d\msd}{dt}\!=\!cD_E.
\end{equation}
This  is just the familiar Green-Kubo expression for normal heat
conduction \cite{Helfand.60.PR,Green.54.JCP,Kubo.57.JPSJ,Luttinger.64.PR,Zwanzig.65.ARPC,Visscher.74.PRA,
Allen.93.PRB}. Arriving at this Green-Kubo relation it is important to recall that in all those cited
derivations one implicitly or explicitly uses the validity of
Fourier's law, together with local thermal equilibrium; i.e. a transport behavior  for steady state heat
flux $j(x)= - \kappa \nabla{T(x)}$. For a small thermal bias
$\Delta T$ the spatially constant gradient scales as $\nabla T(x) = \Delta T/L$.
This in turn implies a length scaling for normal heat conductivity, $\kappa(L)= \kappa
L^{\alpha=0}\equiv \kappa^{normal}$, being independent of system size.
Normal heat diffusion being proportional to time $t$ thus implies with $\beta=1$ the
self-consistent scaling relation, $\alpha = \beta-1=0$.

{\it Superdiffusive energy diffusion.}
With ergodic superdiffusive energy diffusion obeying $\msd\sim t^\expx$,  $1 < \beta \leq
2$, the  time-local Helfand relation (\ref{eq:local_helfand})
possesses no long time limit and
the integral of $\Corr_{JJ}$ diverges as
well. Therefore, {\it no}
finite
superdiffusive
heat conductivity exists.  The typical way out in practice
\cite{Lepri.98.EL,Narayan.02.PRL,Lepri.03.PR,Dhar.08.AP},
however, is to consider a finite system of length $L$ and to formally introduce an upper
cut-off signal time $t_s$ for heat transfer across the sample.  In terms of a characteristic
scale for the speed $v_s$ of phonon transport  one sets $t_s\sim~L/v_s$; $v_s$ is commonly approximated by the speed of sound, being renormalized for nonlinearity \cite{Li.10.PRL}. By adopting this reasoning,
the use of the
time-local Helfand relation (\ref{eq:local_helfand})
{implies then an asymptotic behavior}
\begin{equation}
	\label{eq:kappal}
\kappa_{L}^{{super}}\!\!\sim\!\frac{1}{k_BT^2}\!\int_0^{L/v_s}\!\!\!\Corr_{JJ}(t)d{t}
=\!\frac{
c }{ 2}\!\!
\left.\frac{d\msd}{dt}\right|_{t\sim L/v_s}\!\!\!\!.
\end{equation}
This finite-time Green--Kubo relation implies for the length-dependent
 superdiffusive heat conductivity
$\kappa_L^{{super}}\sim L^\expk$ the scaling relation
\begin{equation}
	\label{eq:ab_new}
	\expk=\expx-1 \;.
\end{equation}
This result corroborates the relation
derived for a specific case of a billiard model where the particles
undergo an {\it a priori} assumed
\levy\ walk process \cite{Denisov.03.PRL,Dhar.13.PRE}.

{\it Subdiffusive energy diffusion.}
Let us next consider an ergodic energy subdiffusion with $\msd\sim t^\expx$, $0<\expx<1$.
From the main relation in (\ref{eq:main})
it follows that the total heat flux correlation $\Corr_{JJ}(t)\sim \expx (\expx-1)
t^{\expx-2}$. With the relation for the exponent, i.e.,  $\delta=\expx-2<-1$, we find
that $\Corr_{JJ}(t)$ remains integrable over the total time
$[\,0,\infty)$.  The time-local   Helfand formula in (\ref{eq:local_helfand}) is  thus
applicable for $t\rightarrow \infty$; yielding
\begin{equation}
\kappa^{{sub}}=\lim_{t\to\infty}\frac{c}{2}\frac{d\msd}{dt}\sim\lim_{
t\to\infty }
t^{\expx-1}=0,
\end{equation}
which indicates a perfect thermal insulator. --- { How does this
vanishing of subdiffusive heat conductivity occur with increasing size $L$?
--- If we likewise may impose in  (\ref{eq:local_helfand}) a finite
cut-off
time scale $t_s \propto L$ we find that ergodic heat subdiffusion occurs with $\kappa^{sub} \sim
L^{\alpha}$, $-1<\alpha=\beta-1<0$.}

{\it Conclusion.} With this work we studied anomalous heat diffusion
in the absence of ergodicity breaking.
The main finding  in (\ref{eq:main})  relates dynamically
the acceleration of the nonequilibrium energy MSD directly
to the equilibrium autocorrelation $\Corr_{JJ}(t)$ of the total heat flux.  Equivalently,
this result assumes  the form of a time-local
Helfand relation as specified with (\ref{eq:local_helfand}).
Given the premise that
anomalous stationary heat flux
follows a behavior in terms of an anomalous heat conductivity, i.e.
$\kappa(L)\sim
L^{\alpha}$,  then implies  the  scaling, $\alpha= \beta-1$. Because
(\ref{eq:main}) applies
for  all times $t$, it  can be invoked  as well for those intermediate cases where an anomalous, length-dependent  heat conductivity
occurs over a finite size
\cite{Zhang.05.JCP,Henry.08.PRL,Henry.09.PRB,Nika.09.APL,Yang.10.NT,
Lindsay.10.PRB, Lindsay.11.PRB, Nika.12.NL,Liu.12.PRB}.

The similarity between  the global Helfand moment scenario used for  normal diffusion in
Ref. \cite{Helfand.60.PR} with the time-local result in (\ref{eq:local_helfand}) suggests
analogous
relations as in (\ref{eq:main}) to hold  for other anomalous diffusion processes.
Particularly, what comes to mind is
unbiased, anomalous particle diffusion $x_p(t)$. Unlike for  energy diffusion
in solid phases, the position increments, i.e., $(x_p(t) -x_p(s)) =\int_s^t {\dot x}_p (t') dt'$,
are now given in terms of the particle velocity $\dot
x_p(t)$. Indeed with ergodic
anomalous diffusion obtained from an equilibrium  GLE-dynamics
\cite{Deng.09.PRE,Goychuk.09.PRE,Goychuk.11.NULL,Siegle.11.EL,Magdziarz.11.APNY,
Sokolov.12.SM}:
with ${\dot x}_p = v(t)$ and $\langle v(t)\rangle_{eq} =0$, $m\langle v^2(t)\rangle_{eq}=
k_BT$, it
readily follows that  (\ref{eq:main})  implies    $d^2\msdp/dt^2 = 2
\langle v(t)v(0)\rangle_{eq}$ for all times $t$ \cite{note}.

\begin{acknowledgments}
This work is supported  by R-144-000-305-112 from MOE T2 (Singapore),
the National
Natural Science Foundation of China, Grant No. 11205114 (N.L.) and the Program for New
Century Excellent Talents of the Ministry of Education of China, Grant No. NCET-12-0409
(N.L.). J.R. acknowledges the support from National Nuclear
Security Administration of the U.S. DOE at LANL under Contract No.
DE-AC52-06NA25396 through the LDRD Program.
\end{acknowledgments}


\begin{thebibliography}{5}
\bibitem{Unit.00.NULL} Following \cite{Li.12.RMP}, we choose the atom
mass $m$, the lattice constant $a$, the force constant $k_0$ and the
Boltzmann constant $k_B$ as the four basic units to scale all
physical quantities involved to dimensionless quantities.
\end{thebibliography}

\clearpage
\onecolumngrid
\setcounter{equation}{0}
\setcounter{page}{1}
\renewcommand{\theequation}{S\arabic{equation}}
\renewcommand{\thepage}{S. \arabic{page}}
\begin{center}
{\large\bf Supplementary Material for ``Anomalous Heat Diffusion''}
\end{center}

\newcommand{\qq}{\{q_i\}}
\newcommand{\pp}{\{p_i\}}
\newcommand{\qqq}{\{q'_i\}}
\newcommand{\ppp}{\{p'_i\}}
\newcommand{\lpartial}[2]{\frac{\partial #1}{\partial #2}}
\newcommand{\poisson}[2]{\left\{#1, #2\right\}}

\newcommand{\meann}[1]{\left\langle#1\right\rangle_{neq}}
\newcommand{\meanE}[1]{\left\langle#1\right\rangle_{E}}
\newcommand{\textcolorr}[2]{#2}
\newcommand{\pt}[2]{\partial_{#2}#1}
\newcommand{\ptt}[3]{\partial_{#3}^2 #2}
\newcommand{\lpt}[2]{\frac{\partial #1}{\partial {#2}}}
\newcommand{\lptt}[3]{\frac{\partial^{#1}{#2}}{\partial {#3}^{#1}}}
\newcommand{\dif}[2]{\frac{\dd#1}{\dd#2}}
\newcommand{\diff}[3]{\frac{\dd^#1 #2}{\dd#3^#1}}

\maketitle
In this supplementary material we detail in a more explicit manner our  theoretical and
numerical analysis used in deriving our main results and provide  additional insight as
needed in our study.

\section{System under study and definitions}
In the following we assume that no particle and charge exchanges assist the energy
transport.
We thus  consider a 1D system given by the Hamiltonian:
\begin{equation}
	H= \sum_n H_n(X),
\end{equation}
where $X$ denotes the complete set of canonical phase space coordinates $(\qq,\pp)$
describing the microscopic system dynamics.
$H(X)$ is composed  as  a sum of the corresponding discrete, local Hamiltonian of the
$n$'th
particle dynamics with the interaction between neighboring particles being  short ranged.
In a space-continuous description this total Hamiltonian then assumes the form
as an integral over a local energy density $h(x)$; i.e.,
\begin{equation}
\label{eqsup:h_continuous}
H = \int h(x) dx, \quad\quad h(x)= \sum_n H_n\delta(x-q_n).
\end{equation}
Given this local energy density  the corresponding local energy current obeys the
condition of local energy conservation,
\begin{equation}
\label{eq:continuity}
\partial_t h(x)+ \partial_x j(x)=0\;,
\end{equation}
or its discrete correspondence. A more detailed discussion and the specific  definitions
in terms of the system parameters and interaction potentials  can be found in
the comprehensive two reviews \cite{Lepri.03.PR,Dhar.08.AP}.

\section{Evolution of the excess energy distribution}
Next, we derive the time-evolution of the excess energy distribution, using  the discrete
version. The corresponding  result  for the space-continuous version follows in a
straightforward manner.

In thermal equilibrium characterized by the  temperature $T$ the probability for the phase
space coordinates obeys with inverse temperature $\beta_T=1/(k_BT)$ the canonical form
\begin{equation}
f_{eq}= \frac{1}{Z} e^{-\beta_T H} \quad\quad \mathrm{with}\quad\quad  Z=\int e^{-\beta_T
H}d\Gamma \;,
\end{equation}
where $d\Gamma=dq_ {1}\cdots dp_{1}\cdots$.
For a prepared nonequilibrium initial phase space probability the time evolution is
governed  by the Liouville equation,
\begin{equation}
\lpartial{}{t}f(t)= L f= \poisson{H}{f}\;,
\end{equation}
where $\poisson{A}{B}$ denotes the Poisson bracket
\begin{equation}
	\poisson{A}{B}=
\sum_i\left(\lpartial{A}{q_i}\lpartial{B}{p_i}-
\lpartial{A}{p_i}\lpartial{B}{q_i}\right).
\end{equation}

Next we introduce a small perturbation $H'$ of the
Hamiltonian, reading:
\begin{equation}
\label{eq:deviation}
H' = - \sum_n \eta_n H_n \;.
\end{equation}
Physically this means that we prepare a nonequilibrium probability, i.e., $f_{neq}(t=0)$, by suddenly switching
off  at $t=0$ the quenched Hamiltonian
$H_T = H + H'$, which is assumed to have acted  since infinite past. Put differently, the
initial-value problem we solve has an initial probability prepared in such a displaced,
frozen-equilibrium ensemble probability, whose future time evolution $f_{neq}(t), t>0$ is
governed by  the unperturbed Liouvillian $L$.  It thus reads
\begin{equation}
\label{eq:ivalue}
f_{neq}(t=0) = \frac{1}{Z'} e^{-\beta_T (H+H')}
\quad\quad \mathrm{with}\quad\quad Z'=\int e^{-\beta_T
(H+H')}d\Gamma \;.
\end{equation}

Using that $H'$ is small, we can expand $Z'$ to linear
order, yielding
\begin{equation}
	Z'= \int e^{-\beta_T H}(1-\beta_T H')d\Gamma=
	Z\left(1-\frac{1}{Z}\int e^{-\beta_T H}\beta_T H'd\Gamma \right)
	= Z(1-\beta_T \mean{H'}).
\end{equation}

As time evolves this nonequilibrium probability for $t>0$ assumes the formal solution
\begin{equation}
\begin{split}
	f_{neq}(t) &= e^{t L}f_{neq}(t=0) = \frac{1}{Z'}e^{t L}e^{-\beta_T H'} e^{-\beta_T H}
\\
	&\approx  \frac{1}{Z}(1+\beta_T \eqi{H'})e^{t L} (1-\beta_T H') e^{-\beta_T H}
	\approx  e^{t L}(1-\beta_T \Delta H') f_{eq} \\
	&= f_{eq}-\beta_T e^{t L} \Delta H' f_{eq},
\end{split}
\end{equation}
where for any quantity $A$, we define $\Delta A= A-\eqi{A}$. The expectation value then
for $H_n(\qq,\pp)$ reads
\begin{equation}
\label{eq:excess_energy}
	{\langle H_n(t)\rangle}_{neq}=\int H_n f_{neq}(t)d \Gamma
	= \eqi{H_n}- \beta_T \int H_n e^{t L} \Delta H' f_{eq}d \Gamma.
\end{equation}

The linear response  in Eq. (\ref{eq:excess_energy}) can  thus be cast in terms of a
stationary equilibrium correlation function of  energy-energy fluctuations, reading
\begin{equation}
	\delta {\langle H_n(t)\rangle}_{neq} = {\langle H_n(t)\rangle}_{neq} - \eqi{H_n(t)}=
-\beta_T \mean{H_n(t) \Delta H'(0)} .
\end{equation}
Using the result in  (\ref{eq:deviation}) we obtain
\begin{equation}
\label{eq:discrete}
	\Delta {\langle H_n(t)\rangle}_{neq}= \sum_i \frac{\eta_i}{k_B T}\mean{\Delta H_n(t)
\Delta H_i(0)}.
\end{equation}

Similarly, the  spatial-continuous version is analogously  given by the initial
nonequilibrium probability density
\begin{equation}
	f_{neq}(t=0)=\frac{1}{Z'} e^{-\beta_T \int [1-\eta(x)]h(x) dx},
\end{equation}
yielding for time evolution of the excess energy density:
\begin{equation}
\label{eq:continuous}
	\delta {\langle h(x,t) \rangle}_{neq}= \frac{1}{k_B T}\int \eta(x')\mean{\Delta h(x,t)
\Delta h(x',0)}dx'.
\end{equation}

Equation (\ref{eq:discrete}) remains valid as well for the system formally connected to
to generalized Langevin heat baths, see in \cite{Grabert.77.ZPB,Grabert.80.JSP}. In such a
case, the Liouville
equation should be replaced by a corresponding, typically non-Markovian, generalized
master equation operator  which determines the evolution of
phase space density. Therefore, the derivation are the same by  replacing the
Liouville operator $L$ with a generalized master operator; i.e.,  $L\to L_{\mathrm{GME}}$
\cite{Grabert.78.PLA}.


\section{Heat capacity and heat-flux autocorrelation function}
In this section, we first demonstrate the  relation
\begin{equation}
\label{eq:SH}
 \lim_{L\to \infty} \int_{-L/2}^{L/2} \Corr_{hh}(x,0)dx=k_BT^2c \;,
\end{equation}
where $c$ denotes the specific volumetric heat capacity. Consider first a continuous
finite system with length $L$ in
thermal equilibrium. Then the total system energy
\begin{equation}
	E_L= \int_{-L/2}^{L/2} h(x,t)dx,
\end{equation}
fluctuates in time.
From a thermal equilibrium statistics,  the variance of this energy fluctuation obeys
\begin{equation}
\label{eq:cL}
	\mean{\Delta E_L\Delta E_L}=k_BT^2 C= k_BT^2 cL\;,
\end{equation}
where $C=cL$ is the total heat capacity for the system of size $L$.

For the spatial correlation of the equilibrium energy density  $\Delta h(x,t)$ we find for
(\ref{eq:cL}) with temporal invariance and observing the fact that this equilibrium
correlation is a
symmetric function of its arguments $(x,x')$, i.e.,
$\Corr_{hh}(x,0;x',0)=\Corr_{hh}(x',0;x,0)$, thus allowing the restriction of integration
to the domain  $x'>x$ by
doubling  the integral:
\begin{equation}
	 \int_{-L/2}^{L/2} dx \int_{-L/2}^{L/2} dx'\mean{\Delta h(x,t)\Delta h(x',t)}=
\int_{-L/2}^{L/2} dx \int_{-L/2}^{L/2} dx'
\Corr_{hh}(x,0;x',0) = 2\int_{-L/2}^{L/2} dx \int_{x}^{L/2} dx' \Corr_{hh}(x,0;x',0).
\end{equation}
We now introduce the difference variable  $y= x'-x$ and use with spatial homogeneity
that $\Corr_{hh}(y,t)= \Corr_{hh}(-y,t)$, followed by a change  of order of integration,
yielding
\begin{equation}
\label{eq:correction}
\begin{split}
2 \int_{-L/2}^{L/2} dx \int_{x}^{L/2}  dx'\; \Corr_{hh}(x,0;x',0)  &= 2
\int_{-L/2}^{L/2}dx \int_{0}^{L/2-x}dy \; \Corr_{hh}(x,0;x+y,0)  \\
&= 2 \int_{-L/2}^{L/2}dx \int_{0}^{L/2-x}dy \;\Corr_{hh}(y,0) \\
&= 2 \int_{0}^{L} dy\;\Corr_{hh}(y,0) \int_{-L/2}^{L/2 -y} dx  \\
&=  2L \int_{0}^{L} dy\; \Corr_{hh}(y,0) \Big(1-\frac{y}{L}\Big)
 .
\end{split}
\end{equation}
For finite time $t$ the integral
$\int_0^\infty \Corr_{hh}(y,t)dy$ must exist. The reasoning goes as
follows.  Because the
spatial-temporal correlation function $\Corr_{hh}(x,t)$ results as  the response to a
sharp perturbation at position $x'=0$ at $t=0$, as shown with
(\ref{eq:continuous}) by considering formally the perturbation $\eta(x')=\delta(x')$. In
physical realistic materials, it always
requires finite time to reach the  cause at  position $x$  due to an applied initial
perturbation at $x=0$; i.e. there is always only a finite speed $v_s$ available  for
information transfer. In our case, this finite speed for information transfer is
characterized by the sound speed $v_s$.
Thus, $\Corr_{hh}(x,t)$ vanishes outside of the causal ``sound cone'', given by $|x|>v_s
t$. This consequently implies
the convergence of $\int_0^\infty \Corr_{hh}(y,0)dy$.
It then follows that for arbitrary finite $t$
\begin{equation}
\lim_{L\to\infty}\int_{0}^{L} dy\;\frac{y}{L} \Corr_{hh}(y,t)=0.
\end{equation}


Noting that  $2 \int_0^{\infty} dy \; \Corr_{hh}(y,0)= \int_{-\infty}^{\infty}dy\;
\Corr_{hh}(y,0)$ and the division in (\ref{eq:cL})  by $L$ we find in this limit of large
system
size  $L$
\begin{equation}
\int_{-\infty}^{\infty} \Corr_{hh}(x,0)dx=\lim_{L\to\infty} \frac{1}{L}\mean{\Delta
E_L\Delta
E_L}=k_BT^2 c\;.
\end{equation}
This shows the  validity of the relation in (\ref{eq:SH}). At best it is only at critical
points with diverging specific volumetric heat capacity $c$ that
$\int_0^\infty \Corr_{hh}(y,0)dy$ may not  converge.

Using the change   $h(x,t)\to$  $j(x,t)$ (the energy current density)  and $E_L\to$  $J_L$
(the the total heat flux),  the same way of reasoning  then yields the result that
\begin{equation}
	C_{JJ}(t)=     \lim_{L\to\infty}\frac{1}{L}\mean{J_L(t)J_L(0)}=
	\int_{-\infty}^{\infty} \Corr_{jj}(x,t)dx\;.
\end{equation}

\section{Relation between energy density correlation and heat flux
density
correlation}
Let us show that
\begin{equation}
\label{eq:hh-jj}
	\frac{\partial^2 \Corr_{hh}(x,t)}{\partial t^2}=
	\frac{\partial^2 \Corr_{jj}(x,t)}{\partial x^2}.
\end{equation}
Using local conservation of energy current we multiply Eq.
(\ref{eq:continuity}) by
$h(x',t')$ and $j(x',t')$ respectively, and take the ensemble
averages:
\begin{eqnarray}
    \label{eq:a1}
\pt{\mean{h(x,t)h(x',t')}}{t}+\pt{{\mean{j(x,t)h(x',t')}}}{x}&=&0,\\
    \label{eq:a2}
\pt{\mean{h(x',t')j(x,t)}}{t'}+\pt{{\mean{j(x',t')j(x,t)}}}{x'}&=&0.
\end{eqnarray}
In the second line, we interchanged  $(x,t)\to(x',t')$.

By performing $\partial_{t'}$ to
Eq.
(\ref{eq:a1}) and $\partial_{x}$ to Eq. (\ref{eq:a2}), we obtain
\begin{equation}
	\frac{\partial^2}{\partial t\partial t'}\mean{h(x,t)h(x',t')}=
	\frac{\partial^2}{\partial x\partial x'}\mean{j(x,t)j(x',t')}
\end{equation}
The time-translational invariance implies that
$\mean{h(x,t)h(x',t')}=\mean{h(x,t-t')h(x',0)}$.  Therefore
\begin{equation}
\label{eq:t-t'}
\frac{\partial^2}{\partial t\partial t'}\mean{h(x,t)h(x',t')}=
-\frac{\partial^2}{\partial t^2}\mean{h(x,t)h(x',t')}.
\end{equation}
For a spatially homogeneous system, this simplifies to yield
$\mean{j(x,t)j(x',t')}=\Corr_{jj}(x-x',t-t')$ so that
\begin{equation}
\label{eq:x-x'}
\frac{\partial^2}{\partial x\partial x'}\mean{j(x,t)j(x',t')}=
-\frac{\partial^2}{\partial x^2}\mean{j(x,t)j(x',t')}\;.
\end{equation}
Observing (\ref{eq:t-t'}) and (\ref{eq:x-x'})  we find  the relation
in (\ref{eq:hh-jj}).

\section{Conservation of excess energy and time independence for mean of energy diffusion}
In this section, we show that for a  homogeneous system, the total excess energy
\begin{equation}
\delta E(t)=\int \delta {\langle h(x,t)\rangle}_{neq}dx=\frac{1}{k_BT}
\iint
\Corr_{hh}(x-x',t)\eta(x')dx'dx,
\end{equation}
 remains conserved. To show this, we take the time derivative twice,
which gives with integration by parts and together with Eq.
\eqref{eq:hh-jj}
 \begin{equation}
	\frac{d^2\delta E(t)}{dt^2}= \frac{1}{k_BT} \iint
\frac{\partial^2\Corr_{hh}(x-x',t)}{\partial t^2}\eta(x')dx'dx
=  \frac{1}{k_BT} \iint
\frac{\partial^2\Corr_{jj}(x-x',t)}{\partial x^2}\eta(x')dx'dx=0 \;.
 \end{equation}
Thus, the first time derivative is a constant.
On the other hand, at $t=0$, we obtain
\begin{equation}
	\frac{d\delta E(0)}{dt}= \frac{1}{k_BT} \iint
\left.\frac{\partial \Corr_{hh}(x-x',t)}{\partial t}\right|_{t=0}\eta(x')dx'dx.
\end{equation}
Note that for any inertial dynamics $\Corr_{hh}(x-x',t)$ is
an even function of $t$, being continuously differentiable at $t=0$. Therefore, the rhs vanishes, yielding
$d\delta E(t)/{dt}$ identically zero, implying that  $\delta
E(t)$ is conserved.

Using a similar reasoning it follows  that the first moment of the excess energy
${\langle x \rangle}_E = \int
x\rho_E(x,t)dx$
remains constant.

\section{Numerical details}
Using dimensionless units \cite{Unit.00.NULL} the Hamiltonian
of the Fermi-Pasta-Ulam (FPU) lattice reads:
\begin{equation}
H=\sum_{i}\left[\frac{1}{2}p^2_i+\frac{1}{2}(q_{i+1}-q_{i})^2+\frac{1}
{4}(q_{i+1}-q_{i}
)^4\right] \;.
\end{equation}
Here, the set ${q_i}$ denotes the relative displacement with respect to
the equilibrium position $ia$ and ${p_i}$ denotes the momentum
for the $i$-th atom, where $a$ is the lattice constant
which can be scaled to unity, i.e., $a=1$ \cite{Unit.00.NULL}. We
further use periodic
boundary conditions; i.e., $q_{N+1}=q_1$.
The lattice length is $L=Na$ with $N=401$. The local
energy $H_i(t)$ is then chosen as:
\begin{equation}
H_i(t)=\frac{1}{2}p^2_i+\frac{1}{2}\big[
V(q_{i}-q_{i-1})+V(q_{i+1}-q_{i})\big];\quad V(x)=
\frac{1}{2}x^2+\frac{1}
{4}x^4.
\end{equation}
For convenience, the atom indexes are chosen as
$i=-200,\cdots,200$. In the simulation, the dimensionless time step
size is
set to $\tau=0.05$.

\begin{figure}[t]
\includegraphics[width=0.65\columnwidth]{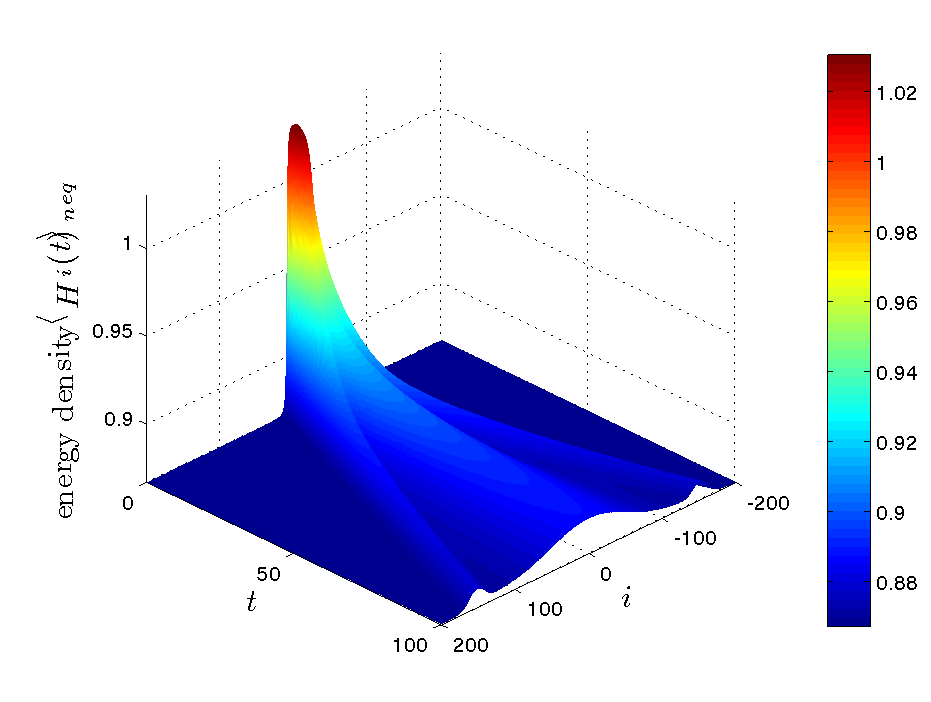}
\caption{(Color online) The time evolution of the nonequilibrium
energy density  for a manifest
near equilibrium energy diffusion dynamics.}
\label{fig:sup}
\end{figure}

To evaluate  both,  $C_{hh}(x,t)$ in  linear response,   Eq.
(\ref{eq:discrete}), and the heat flux autocorrelation function
$C_{JJ}(t)$ in thermal equilibrium,
we first apply Langevin heat baths at temperature $T=1$ to all
atoms.   The velocity-Verlet algorithm
is used. Doing so does prepare the canonical equilibrium state.
After all transients have died out, the heat
baths are removed. Then a fourth order symplectic SABA${}_2$C
algorithm \cite{Laskar.01.CMDA} is used to integrate the equations of
motion and the corresponding correlation functions are calculated. The final
correlation function is based on an average over $2\times 10^9$
realizations.
For our illustration in Fig. 1(a), the excess energy distribution are
based on Eq. (\ref{eq:discrete}), using an initial excess energy
profile $\eta_i$, being composed of
two Gaussian peaks, one with positive and one with negative weight; i.e. we set:
\begin{equation}
\label{eq:T}
	\eta_i= 10^{-3}\;
\Big[\exp{\left(-\frac{(i-20)^2}{2\times12^2}\right)}-
\exp{\left(-\frac{(i+30)^2}{2\times8^2}\right)}\Big]\;.
\end{equation}

To simulate a full nonequilibrium energy diffusion, we first prepare
the system in a nonequilibrium steady
state near a reference temperature $T=1$. Specifically, we apply
Langevin heat baths to all atoms with different temperatures:
\begin{equation}
	T_i= \begin{cases}
	       1.2 & \mathrm{for }\ -10\le i\le 10; \\
	       1.0 & \mathrm{otherwise}.
	     \end{cases}
\end{equation}

%
%
We use velocity-Verlet algorithm and run for
$1\times 10^7$ steps to reach the nonequilibrium steady state. Then all the heat
baths are removed and the energy profiles are calculated up to
time $t=100$ using the fourth order symplectic SABA${}_2$C algorithm.
An ensemble of $4\times 10^8$ realizations are used to evaluate the time evolution of the
nonequilibrium energy density $\meann{H_i(t)}$ as depicted in Fig.
(\ref{fig:sup}). The normalized  energy distribution
$\rho_E(x,t)$ is calculated using
\begin{equation}
 \rho_E(x=i,t)= \frac{\meann{H_i(t)}- \mean{H_i}}{\sum_i
\left[\meann{H_i(t)}- \mean{H_i}\right]},
\end{equation}
where the reference energy density $\mean{H_i}$ is set to
the average energy density at reference temperature $T=1$, which
equals $0.867$, see in Eq. \eqref{eq:e} below.

Finally, the MSD is calculated using Eq. (3) in the main article
and the second time derivate is calculated using the formula
\begin{equation}
	\frac{d^2 f(t)}{d t^2}= \frac{f(t+\Delta t)-2f(t)+f(t-\Delta
t)}{\Delta t^2}
\end{equation}
with $\Delta t=20h=1$.

The volumetric specific heat $c$ is calculated analytically according
to its definition
\begin{equation}
	c=\frac{d \mean{H_i(T)}}{d T},
\end{equation}
where $\mean{H_i(T)}$ is the average energy per particle at
temperature $T$, which can be calculated as \cite{Unit.00.NULL}
\begin{equation}
\label{eq:e}
	\mean{H_i(T)}= \mean{e_{kinetic}}+\mean{e_{potential}}=
\frac{1}{2}T+ \frac{\int V(x) e^{-V(x)/T}dx}{\int e^{-V(x)/T}dx}.
\end{equation}
For $T=1$, we obtain $\mean{H_i(T)}=0.867$ and $c=0.828$.

\end{document}